\documentclass[aps,prd,reprint,superscriptaddress,nofootinbib,floatfix]{revtex4-2}

\usepackage{amsmath,amssymb,bm}
\usepackage{graphicx}
\usepackage{booktabs}
\usepackage{xcolor}
\usepackage{hyperref}
\usepackage{slashed}

\hypersetup{
  colorlinks=true,
  linkcolor=blue!55!black,
  citecolor=blue!55!black,
  urlcolor=blue!55!black
}

\newcommand{\psitwos}{\psi(3686)}
\newcommand{\OO}{\Omega^-\bar\Omega^+}
\newcommand{\Neg}{\mathcal N}
\newcommand{\SN}{\mathrm{SN}}
\newcommand{\Tr}{\operatorname{Tr}}

\begin{document}

\title{\texorpdfstring{Beam-polarization-enabled witness of
 four-dimensional entanglement in $\Omega^-\bar\Omega^+$ pairs}
 {Beam-polarization-enabled witness of four-dimensional entanglement in
 Omega-anti-Omega pairs}}

\author{Qiu-Yan Zhang}
\email{zhangqy@ihep.ac.cn}
\affiliation{College of Physics, Jilin University, Changchun 130012, China}

\author{Peng-Cheng Hong}
\email{hongpc@ihep.ac.cn}
\affiliation{College of Physics, Jilin University, Changchun 130012, China}
\affiliation{Institute of High Energy Physics, Chinese Academy of Sciences,
Beijing 100049, China}

\author{Wei-Min Song}
\email{weiminsong@jlu.edu.cn}
\affiliation{College of Physics, Jilin University, Changchun 130012, China}

\date{\today}

\begin{abstract}
We study the entanglement structure of $\Omega^-\bar{\Omega}^+$ pairs produced in $e^+e^-\to\psi(3686)$, focusing on whether the entanglement requires the full four-dimensional spin space of each spin-$3/2$ baryon. 
Beam polarization controls the parent-state populations and coherence, thereby modifying helicity interference.
Using the negativity criterion $\Neg>1$ and the measured BESIII helicity parameters, we find critical polarizations of about $0.65$ per beam for equal transverse polarization and $0.57$ per beam for equal-and-opposite longitudinal polarization.  Full transverse polarization gives a
near-maximal peak negativity, $\Neg_{\rm peak}\simeq1.50$, with a production-rate-weighted
angular fraction of $0.60$ satisfying the witness.  At the optimal
production directions, destructive interference removes the unequal-helicity
channels, while the
near equality of the relevant equal-helicity amplitudes yields nearly equal
Schmidt coefficients.  Constituent-quark benchmarks further show that the
angular extent of the witness is more sensitive to relative phases than the peak
negativity.  These results link polarization-controlled helicity
interference to four-dimensional entanglement in high-spin baryon
production.
\end{abstract}

\maketitle

\section{Introduction}
\label{sec:introduction}

Vector charmonium decays into baryon--antibaryon pairs encode their spin
structure in complex helicity amplitudes, whose magnitudes and relative phases determine the baryon polarizations and joint spin correlations \cite{Perotti2019}. The connection between helicity interference and measurable polarization is illustrated by the BESIII determination of the relative production phase in $J/\psi\to\Lambda\bar\Lambda$ \cite{BESIIILambda2019}. Spin-$3/2$ baryon pairs offer a richer helicity structure. In $\psitwos\to\OO$, the two baryons each possess a four-dimensional local spin space, while four independent complex helicity amplitudes characterize the allowed production channels within the
$4\otimes4$ spin space \cite{Perotti2019}. The recent BESIII joint angular analysis has measured the corresponding amplitude ratios and relative phases \cite{BESIIIOmega2026}, providing experimental input for the full two-baryon spin density matrix. This opens the way to investigating how the interplay of helicity channels shapes the quantum correlations of a high-spin baryon pair.

Entanglement measures and Bell observables provide tools for characterizing the quantum correlations encoded in these joint spin states. Their experimental relevance has been demonstrated in top--antitop production at the LHC \cite{ATLAS2024,CMS2024} and in hyperon--antihyperon systems at BESIII, where spin correlations have been exploited for joint measurements of production and weak-decay parameters, probes of $CP$ symmetry and weak phases, and tests of local realism \cite{BESIIILambda2019,BESIIIXi2022,BESIIIBell2025}. Theoretical studies connect charmonium production amplitudes to spin density matrices, entanglement measures, and Bell observables
\cite{Fabbrichesi2024,Wu2024,Hong2026}. The dimensionality of these quantum correlations becomes particularly relevant for higher-spin systems: spin-$1/2$ pairs occupy a $2\otimes2$ spin space, whereas spin-$3/2$ pairs access $4\otimes4$ \cite{Perotti2019}, allowing their entanglement
dimensionality to be characterized through the Schmidt number \cite{TerhalHorodecki2000}.

Entanglement in the unpolarized $\psitwos\to\OO$ channel has already been established using measured helicity amplitudes \cite{Fabbrichesi2024}. The question addressed here concerns the dimensionality of that entanglement. Entanglement in a $4\otimes4$ system does not by itself imply Schmidt number four: an entangled state may have Schmidt number two or three. Certifying Schmidt number four establishes that every pure-state decomposition requires at least one component of Schmidt rank four \cite{TerhalHorodecki2000}. This stronger criterion therefore tests whether the production dynamics generates entanglement that genuinely requires the full local spin dimension. The central question is under what conditions
such four-dimensional entanglement can be certified in
$\psitwos\to\OO$ production.

Beam polarization provides a direct handle on this question by controlling the spin state of the parent charmonium. 
Longitudinal polarization changes the relative populations of the \(m=\pm1\) spin-projection components of the parent charmonium along the beam axis, while transverse polarization of both beams generates coherence between these components.
Through the decay amplitudes, these two controls modify the weights and interference of the final-state helicity channels. Polarized beams have already been considered as tools for hyperon $CP$ tests at future $e^+e^-$ facilities \cite{Cao2024} and for decay-parameter measurements at the Super Tau-Charm Facility (STCF) \cite{Zeng2023}, and their effects on entanglement and Bell nonlocality have also been studied for spin-$1/2$ hyperon pairs \cite{Zhang2026}. Beam polarization therefore provides a
means of reshaping the helicity interference that governs the dimensionality of the final-state entanglement in the richer $\Omega^-\bar\Omega^+$ system.

Using the measured BESIII helicity amplitudes, we construct the polarized $\Omega^-\bar\Omega^+$ spin density matrix and apply the negativity criterion $\Neg>1$ to witness Schmidt number four. We determine the critical polarizations for pure longitudinal and transverse configurations, examine their combined effect over the physical polarization domain, and quantify the production-rate-weighted angular extent of the witness.
Correlated helicity-parameter uncertainties are propagated to the entanglement observables to assess the robustness of the certification. To disentangle the roles of helicity magnitudes and relative phases, we further compare the measured amplitudes with an illustrative constituent-quark benchmark with controlled phase assignments, thereby probing how the amplitude pattern and helicity interference shape the entanglement.

\section{Formalism}
\label{sec:formalism}

\subsection{Helicity amplitudes}
\label{subsec:helicity}

We consider $e^+e^-\to\psitwos\to
\Omega^-(\lambda_1)\bar\Omega^+(\lambda_2)$
in the $e^+e^-$ center-of-mass frame. The $z$ axis follows the incoming
positron direction, consistent with the convention of the BESIII
analysis \cite{BESIIIOmega2026}. The $\Omega^-$ momentum direction is
specified by the Euler angles $(\phi,\theta,0)$, and
$\lambda_{1,2}=\pm\frac12,\pm\frac32$ denote the baryon helicities.
Below, $d\Omega\equiv d\phi\,d(\cos\theta)$ denotes the production
solid-angle element.

In the Jacob--Wick convention \cite{JacobWick1959}, the production amplitude
for a vector-state spin projection $m=0,\pm1$ along the $z$ axis is
\begin{equation}
 \mathcal M_{m;\lambda_1\lambda_2}(\theta,\phi)
 =\sqrt{\frac{3}{4\pi}}
 D^{1*}_{m,\lambda_1-\lambda_2}(\phi,\theta,0)
 A_{\lambda_1\lambda_2},
 \label{eq:helicity-amplitude}
\end{equation}
where $D^1_{m,\lambda_1-\lambda_2}$ is the Wigner $D$ matrix and
$A_{\lambda_1\lambda_2}$ denotes the helicity amplitude. Angular-momentum
conservation requires $|\lambda_1-\lambda_2|\leq1$. For the ground-state
$\Omega^-\bar\Omega^+$ pair, parity conservation and charge-conjugation
invariance imply, in the helicity-frame phase convention adopted here
\cite{Perotti2019},
\begin{align}
 A_{\lambda_1\lambda_2}
 &=A_{-\lambda_1,-\lambda_2}, &&(P),
 \label{eq:parity-helicity}\\
 A_{\lambda_1\lambda_2}
 &=A_{\lambda_2\lambda_1}, &&(C).
 \label{eq:charge-helicity}
\end{align}
In the second relation, the minus sign from the $C=-1$ vector state is
compensated by reordering the two anticommuting fermion fields. Thus the ten
helicity combinations allowed by $|\lambda_1-\lambda_2|\leq1$ are reduced
to five by parity and then to four independent complex amplitudes by charge
conjugation. A convenient independent set is
$A_{1/2,-1/2}$, $A_{1/2,1/2}$, $A_{3/2,1/2}$, and $A_{3/2,3/2}$,
with the remaining amplitudes fixed by
Eqs.~\eqref{eq:parity-helicity} and \eqref{eq:charge-helicity}.

To permit direct comparison with the experimental fit, we adopt the
helicity-parameter definitions of the BESIII analysis
\cite{BESIIIOmega2026}. Taking $A_{1/2,-1/2}$ to be real and positive,
we define
\begin{align}
 \frac{A_{1/2,1/2}}{A_{1/2,-1/2}}&=h_1e^{i\phi_1},
 &
 \frac{A_{3/2,1/2}}{A_{1/2,-1/2}}&=h_3e^{i\phi_3},
 \nonumber\\
 \frac{A_{3/2,3/2}}{A_{1/2,-1/2}}&=h_4e^{i\phi_4}.
 \label{eq:helicity-ratios}
\end{align}
Here $h_i$ denote the amplitude ratios and $\phi_i$ the corresponding
relative phases. The overall normalization and common phase cancel from
the normalized spin density matrix. The helicity-amplitude input is
therefore specified by six real parameters
$(h_1,\phi_1,h_3,\phi_3,h_4,\phi_4)$.

\subsection{Polarized vector-state density matrix}
\label{subsec:initial-density}

Neglecting the electron mass, the directly produced $\psitwos$ has no
$m=0$ component, although its full spin space is three-dimensional.
We denote the transverse polarization magnitudes of the electron and
positron beams by $P_T^-$ and $P_T^+$, respectively, and take
$P_T^-=P_T^+\equiv P_t$. For the longitudinal polarization components,
denoted by $P_z$ and $P_{\bar z}$ for the electron and positron beams,
respectively, we take
$P_z=P_L$ and $P_{\bar z}=-P_L$. Here $P_t\geq0$ and $P_L\geq0$
denote the corresponding polarization magnitudes, and the signs follow
our helicity-basis convention.

With these conventions, the normalized $\psitwos$ density matrix obtained
from the polarized $e^+e^-$ initial state, in the ordered basis
$(m=+1,0,-1)$, is
\begin{equation}
 \rho_\psi(P_t,P_L)=\frac{1}{2(1+P_L^2)}
 \begin{pmatrix}
 (1-P_L)^2 & 0 & P_t^2 e^{-i\delta_T}\\
 0 & 0 & 0\\
 P_t^2 e^{i\delta_T} & 0 & (1+P_L)^2
 \end{pmatrix}.
 \label{eq:initial-normalized}
\end{equation}
Here $\delta_T$ is the relative transverse phase \cite{Cao2024}. Before
setting the two transverse magnitudes equal, the off-diagonal coherence is
proportional to $P_T^-P_T^+$ and therefore vanishes if either beam is
transversely unpolarized. We choose the transverse axes such that
$\delta_T=0$; any other fixed value merely shifts the azimuthal origin.
For each beam, the transverse and longitudinal polarization components obey
the physical constraint
\begin{equation}
 P_t^2+P_L^2\leq1.
 \label{eq:physical-domain}
\end{equation}

For comparison with a configuration in which only the electron beam is
longitudinally polarized, we denote its longitudinal polarization by $P_e$.
The same relative population asymmetry of the $m=\pm1$ parent states is
then obtained for
\begin{equation}
 |P_e|=\frac{2P_L}{1+P_L^2},
 \label{eq:single-beam-mapping}
\end{equation}
where $P_L$ denotes the equal-and-opposite longitudinal polarization
magnitude used in Eq.~\eqref{eq:initial-normalized}. This relation is used
below only to translate the longitudinal witness threshold between the two
beam-polarization conventions.

\subsection{\texorpdfstring{$\Omega^-\bar\Omega^+$}{Omega-anti-Omega}
spin density matrix}
\label{subsec:pair-density}

The unnormalized pair density matrix is
\begin{align}
 R_{\lambda_1\lambda_2;\lambda'_1\lambda'_2}
 (\theta,\phi;P_t,P_L)
 ={}&\sum_{m,m'=\pm1}(\rho_\psi)_{mm'}
 \mathcal M_{m;\lambda_1\lambda_2}(\theta,\phi)
 \nonumber\\[-2pt]
 &\times
 \mathcal M^*_{m';\lambda'_1\lambda'_2}(\theta,\phi).
 \label{eq:pair-density-raw}
\end{align}

At fixed production angles and beam polarizations, the normalized pair spin
state is
\begin{equation}
 \rho(\theta,\phi;P_t,P_L)
 =\frac{R(\theta,\phi;P_t,P_L)}
 {\Tr R(\theta,\phi;P_t,P_L)}.
 \label{eq:pair-density-normalized}
\end{equation}
The trace $\Tr R(\theta,\phi;P_t,P_L)$ is proportional to the differential
production rate at the corresponding production angle. The normalized
state acts on the $4\otimes4$ spin space and is determined by the six
helicity parameters and the beam polarizations. Angular-momentum
conservation restricts its support to the ten joint helicity states
satisfying $|\lambda_1-\lambda_2|\leq1$.

The off-diagonal $m=\pm1$ coherence in
Eq.~\eqref{eq:initial-normalized} propagates through the decay amplitudes
into the pair density matrix. Together with the azimuthal phases of the
Wigner $D$ matrices in Eq.~\eqref{eq:helicity-amplitude}, it generates
terms proportional to $e^{\pm2i\phi}$ in
Eq.~\eqref{eq:pair-density-raw}, leading to a $\pi$-periodic azimuthal
modulation. For pure longitudinal polarization, the off-diagonal coherence
vanishes and the negativity has no physical $\phi$ dependence.

\subsection{Negativity and entanglement dimensionality}
\label{subsec:negativity}

The partial transpose with respect to the antibaryon spin space is defined by
\begin{equation}
 \left(\rho^{T_{\bar\Omega}}\right)_{ij;kl}
 =\rho_{il;kj},
 \label{eq:partial-transpose}
\end{equation}
where $i,k$ label the $\Omega^-$ spin states and $j,l$ label the
$\bar\Omega^+$ spin states. We quantify entanglement using the negativity
\cite{VidalWerner2002},
\begin{equation}
 \Neg(\rho)=\frac{\lVert\rho^{T_{\bar\Omega}}\rVert_1-1}{2}
 =-\sum_{\nu_a<0}\nu_a,
 \label{eq:negativity}
\end{equation}
where $\lVert X\rVert_1=\Tr\sqrt{X^\dagger X}$ denotes the trace norm and
$\nu_a$ are the eigenvalues of $\rho^{T_{\bar\Omega}}$. For a
$4\otimes4$ state,
\begin{equation}
 0\leq\Neg\leq\frac{3}{2}.
\end{equation}

For mixed states, the Schmidt number, denoted by $\SN(\rho)$, characterizes
the entanglement dimensionality. It is the minimum, over all pure-state
decompositions, of the largest Schmidt rank that appears
\cite{TerhalHorodecki2000}. For a pure state of Schmidt rank at most $k$,
let $\lambda_i$ denote the squared Schmidt coefficients, with
$\sum_i\lambda_i=1$. Its negativity satisfies
\[
 \Neg
 =\frac{\left(\sum_i\sqrt{\lambda_i}\right)^2-1}{2}
 \leq\frac{k-1}{2},
\]
where the upper bound follows from the Cauchy--Schwarz inequality.
Convexity of the negativity then extends this bound to mixed states:
\begin{equation}
 \SN(\rho)\leq k
 \quad\Longrightarrow\quad
 \Neg(\rho)\leq\frac{k-1}{2}.
 \label{eq:schmidt-neg-bound}
\end{equation}
Consequently, for the $4\otimes4$ spin space,
\begin{equation}
 \Neg>\frac12\Rightarrow\SN(\rho)\geq3,
 \qquad
 \Neg>1\Rightarrow\SN(\rho)=4.
 \label{eq:dimensional-witness}
\end{equation}
These implications provide sufficient dimensionality witnesses; their
converses do not hold. In particular, $\Neg\leq1$ does not exclude
Schmidt number four.

\section{Numerical setup}
\label{sec:numerical-setup}

\subsection{BESIII helicity inputs}
\label{subsec:besiii-input}

We use the BESIII measurements of the helicity-amplitude parameters reported in
Ref.~\cite{BESIIIOmega2026}.
For the numerical calculations, we use the unrounded central values,
statistical uncertainties, and the six-parameter statistical correlation
matrix corresponding to the final fit.  Calculations retain the full
numerical precision, while Table~\ref{tab:besiii-input} displays two decimal
places.  The first uncertainty is statistical and the second systematic.

\begin{table}[t]
\caption{BESIII measurements of the helicity parameters for $\psitwos\to\OO$
reported in Ref.~\cite{BESIIIOmega2026}.  Phases are in radians.}
\label{tab:besiii-input}
\begin{ruledtabular}
\begin{tabular}{c@{\qquad}c}
Parameter & Fit result\\
\hline
$h_1$    & $0.65\pm0.05\pm0.03$\\
$\phi_1$ & $0.86\pm0.11\pm0.05$\\
$h_3$    & $0.49\pm0.05\pm0.02$\\
$\phi_3$ & $2.18\pm0.07\pm0.04$\\
$h_4$    & $0.66\pm0.03\pm0.01$\\
$\phi_4$ & $3.27\pm0.11\pm0.05$\\
\end{tabular}
\end{ruledtabular}
\end{table}

\subsection{Constituent-quark amplitude benchmark}
\label{subsec:quark-input}

For an illustrative comparison of helicity-amplitude patterns, we
construct a simplified constituent-quark benchmark following the
three-gluon helicity-amplitude framework of Ref.~\cite{PangPing2007}.
The constituents are projected onto the symmetric spin-$3/2$ wave
functions of the $\Omega^-$ and $\bar\Omega^+$.  In the collinear,
equal-sharing approximation, each quark carries one third of its baryon's
momentum and energy,
\begin{equation}
 m_q=\frac{M_\Omega}{3},
 \qquad
 E_q=\frac{E_\Omega}{3},
\end{equation}
where $m_q$ and $E_q$ denote the constituent-quark mass and energy,
respectively, while $M_\Omega$ and $E_\Omega$ denote the $\Omega$ mass and
energy in the $e^+e^-$ center-of-mass frame.

Combining the quark currents with these spin wave functions and
projecting onto the baryon helicity states gives four independent helicity amplitudes, denoted by $F_{\lambda_1,\lambda_2}$.  Their derivation is
summarized in Appendix~\ref{app:quark-amplitudes}.  The same angular-momentum
and discrete-symmetry
constraints as in Sec.~\ref{subsec:helicity} reduce the model to four
independent channels.  Up to a common color, coupling, spatial-wave-function,
and normalization factor, the amplitudes are
\begin{align}
 F_{3/2,3/2}&=\frac{8}{9}M_\Omega^3,
 \nonumber\\
 F_{3/2,1/2}&=\frac{8}{9}\sqrt{\frac{2}{3}}E_\Omega M_\Omega^2,
 \nonumber\\
 F_{1/2,1/2}&=\frac{8}{27}
 \left(-4E_\Omega^2M_\Omega+3M_\Omega^3\right),
 \nonumber\\
 F_{1/2,-1/2}&=-\frac{16\sqrt2}{27}E_\Omega
 \left(2E_\Omega^2-M_\Omega^2\right).
 \label{eq:quark-amplitudes}
\end{align}
Here
\begin{equation}
 E_\Omega=\frac{m_{\psi(3686)}}{2}.
\end{equation}
Within this benchmark, the common factor cancels in the helicity-amplitude
ratios.

Using the mass inputs
$m_{\psi(3686)}=3.686\ \mathrm{GeV}$ and
$M_\Omega=1.6724\ \mathrm{GeV}$, and normalizing the model magnitudes to
$|F_{1/2,-1/2}|$, gives
\begin{equation}
 (h_1,h_3,h_4)=(0.42,0.61,0.67).
 \label{eq:quark-ratios}
\end{equation}
These real amplitudes determine the relative magnitude pattern used for
comparison with BESIII.  Their signs depend on the external-state phase
convention, while physical relative phases may also receive contributions
from final-state rescattering.  Using the same reference-phase convention
as in Sec.~\ref{subsec:helicity}, we consider two phase benchmarks,
\begin{align}
 {\cal Q}_0 &: (\phi_1,\phi_3,\phi_4)=(0,0,0),
 \nonumber\\
 {\cal Q}_{\pi/2} &: (\phi_1,\phi_3,\phi_4)
 =(\pi/2,\pi/2,\pi/2).
 \label{eq:quark-phase-benchmarks}
\end{align}
These choices keep the helicity magnitudes fixed while probing how a common
$\pi/2$ shift of the three relative phases modifies the helicity
interference.

\subsection{Scan observables}
\label{subsec:scan-observables}

For fixed beam polarization, we define the angular extrema
\begin{align}
 \Neg_{\rm peak}(P_t,P_L)
 &=
 \max_{\substack{-1\leq\cos\theta\leq1\\
                 0\leq\phi\leq2\pi}}
 \Neg(\theta,\phi;P_t,P_L),
 \label{eq:npeak}\\
 \Neg_{\rm min}(P_t,P_L)
 &=
 \min_{\substack{-1\leq\cos\theta\leq1\\
                 0\leq\phi\leq2\pi}}
 \Neg(\theta,\phi;P_t,P_L).
 \label{eq:nmin}
\end{align}
The peak determines whether there exists at least one production direction
for which the four-dimensional witness is satisfied, whereas the angular
minimum characterizes the least favorable production direction.

For the pure transverse and pure longitudinal configurations, respectively,
we define the critical polarization magnitudes by
\begin{align}
 P_t^{\rm crit}
 &=
 \inf\left\{
 P_t\in[0,1]:
 \Neg_{\rm peak}(P_t,0)>1
 \right\},
 \label{eq:ptcrit}\\
 P_L^{\rm crit}
 &=
 \inf\left\{
 P_L\in[0,1]:
 \Neg_{\rm peak}(0,P_L)>1
 \right\}.
 \label{eq:plcrit}
\end{align}
If the witness is not reached up to unit polarization, no crossing is
assigned.  The critical polarization is obtained by first bracketing the
crossing of $\Neg_{\rm peak}-1$ on a polarization grid with spacing $0.1$
and then refining the crossing with Brent's method.  The angular maximum is
reoptimized at every trial polarization.

To quantify the angular extent of the pointwise four-dimensional witness,
we use the production-rate-weighted fraction
\begin{widetext}
\begin{equation}
f_{\Neg>1}(P_t,P_L)
=
\frac{
\displaystyle
\int d\Omega\,
\Tr R(\theta,\phi;P_t,P_L)\,
\Theta\!\left[\Neg(\theta,\phi;P_t,P_L)-1\right]
}{
\displaystyle
\int d\Omega\,
\Tr R(\theta,\phi;P_t,P_L)
} ,
\label{eq:weighted-witness-fraction}
\end{equation}
\end{widetext}
where $\Theta$ is the Heaviside step function.  When discussing optimal
polarizations below, we distinguish between configurations that maximize
$\Neg_{\rm peak}$ and those that maximize $\Neg_{\rm min}$.  For the
combined polarization optimization, we parameterize
$P_t=r\cos\beta$ and $P_L=r\sin\beta$, with $0\leq r\leq1$ and
$0\leq\beta\leq\pi/2$.  Candidate extrema are first located by scans
over the physical polarization domain and then refined by continuous
optimization, with the production-angle extrema reoptimized at every
polarization point.  The continuously refined optima are used for the
quoted results; for the central BESIII inputs, successively finer
polarization-domain scans and independent continuous optimization reproduce
the same global optimum.

We compare the BESIII inputs and the two quark-model phase benchmarks over
the full physical polarization domain in Eq.~\eqref{eq:physical-domain}
and over all production angles.  The critical polarizations, optimized
angular extrema, and production-rate-weighted witness fractions
characterize how the amplitude pattern and its interference shape the
entanglement.

For the central BESIII inputs, refining the
$(\cos\theta,\phi)$ grid from $61\times73$ to $121\times145$ and
$241\times289$ points changes the two critical polarizations, the
full-transverse peak, and its angular minimum by less than $10^{-6}$;
the optimal angles are unchanged up to symmetry.  The weighted witness
fraction changes by less than $10^{-4}$ when the
$(\cos\theta,\phi)$ integration grid is refined from $401\times721$
to $801\times1441$.  These checks support the two-decimal precision of
the quoted observables.

\subsection{Uncertainty treatment}
\label{subsec:uncertainty}
We reconstruct the statistical covariance matrix from the unrounded statistical uncertainties and the six-parameter statistical correlation matrix described in Sec.~III A.  Because a systematic correlation matrix
is not available, the systematic uncertainties reported in
Ref.~\cite{BESIIIOmega2026} are added in quadrature to the covariance diagonal, corresponding to the approximation that the systematic shifts
of the six parameters are mutually independent.  The total covariance matrix used for sampling is therefore
\begin{equation}
 C_{\rm tot}
 =
 C_{\rm stat}
 +
 \operatorname{diag}
 \left(
 \sigma_{\rm syst}^2
 \right).
 \label{eq:total-covariance}
\end{equation}
Here $C_{\rm stat}$ is the statistical covariance matrix, and
$\sigma_{\rm syst}$ denotes the systematic uncertainties of the six
helicity parameters.

We generate 10000 correlated Gaussian samples of the six helicity parameters
using $C_{\rm tot}$.  Samples with any negative amplitude magnitude are
rejected and redrawn, while the phases are wrapped modulo $2\pi$.  For each
sampled parameter set, we determine both polarization thresholds and the
full-transverse peak and weighted angular extent.  At each polarization
point, the production-angle extrema are independently recomputed.

At the full-transverse point $P_t=1$ and $P_L=0$, we define the pointwise
sample fraction retaining the four-dimensional witness as
\begin{equation}
 f_{\rm samp}(\Neg>1;\theta,\phi)
 =
 \frac{1}{N_{\rm samp}}
 \sum_{i=1}^{N_{\rm samp}}
 \mathbf{1}\!\left[
 \Neg_i(\theta,\phi)>1
 \right].
 \label{eq:sample-witness-fraction}
\end{equation}
Here $N_{\rm samp}=10000$, $\Neg_i(\theta,\phi)$ denotes the negativity
evaluated for the $i$th sampled helicity-parameter set, and
$\mathbf{1}[\cdots]$ is the indicator function.  To quantify the
angular region that retains the witness for at least 95\% of the sampled
parameter sets, we use the production rate evaluated with the central
BESIII helicity parameters,
$w_{\rm c}(\theta,\phi)\equiv
\Tr R(\theta,\phi;P_t=1,P_L=0)$, and define
\begin{equation}
 f_{95}^{(\rm c)}
 =
 \frac{
 \displaystyle
 \int d\Omega\,
 w_{\rm c}(\theta,\phi)\,
 \mathbf{1}\!\left[
 f_{\rm samp}(\Neg>1;\theta,\phi)\geq0.95
 \right]
 }{
 \displaystyle
 \int d\Omega\,
 w_{\rm c}(\theta,\phi)
 }.
 \label{eq:robust-witness-fraction}
\end{equation}
The superscript $(\rm c)$ denotes this central-input production-rate
weighting.  The central-input rate provides a common angular weighting for
comparing the pointwise robustness across parameter samples.  By contrast,
each sample-specific value of $f_{\Neg>1}$ is evaluated using the production
rate of that same parameter sample.  The
angular integrals in Eq.~\eqref{eq:robust-witness-fraction} are evaluated
with trapezoidal endpoint weights.  Using the original 1000-sample validation
ensemble, refining the $(\cos\theta,\phi)$ grid from $101\times181$ to
$201\times361$ changes $f_{95}^{(\rm c)}$ by $4.1\times10^{-4}$, leaving
its quoted two-decimal value unchanged.

Medians and central 68\% and 95\% intervals characterize the propagated
variation within this uncertainty model.  Threshold intervals are
conditional on a crossing; samples that remain below the witness threshold
at $P_t=1$ or $P_L=1$, respectively, are counted separately.  Increasing
the correlated parameter ensemble from 5000 to 10000 samples leaves the
quoted longitudinal median and central intervals unchanged at the reported
precision; the longitudinal no-crossing fraction changes from 5.1\% to
5.0\%.  The transverse-threshold results, the witness-fraction and
peak-negativity intervals, and $f_{95}^{(\rm c)}$ also remain unchanged at
the quoted precision.  We therefore use the 10000-sample ensemble for the
final uncertainty results.

As a numerical stability check, we retain the original 1000-sample ensemble
as a nested validation subset of the final 10000-sample calculation.  Within this
validation subset, we repeat the threshold calculation on a refined angular grid
for a deliberately selected set of 168 samples.  This set comprises 100
samples uniformly spaced over the original 1000-sample validation ensemble, all
samples without a longitudinal crossing in that ensemble, and the five
smallest and five largest crossing thresholds for each of the transverse
and longitudinal configurations, with duplicates removed.  Refining the
$(\cos\theta,\phi)$ grid from $41\times49$ to $81\times97$ changes no
crossing classification and alters the extracted thresholds only at
numerical precision.

\section{Numerical results}
\label{sec:results}

\subsection{Unpolarized benchmark}

For $P_t=P_L=0$, the negativity is independent of $\phi$ and symmetric under
$\cos\theta\to-\cos\theta$.  For the central BESIII inputs, the angular
maximum is
\[
 \Neg_{\rm peak}(0,0)=0.73,
\]
attained at $\cos\theta=0$, while the full-domain minimum is
\[
 \Neg_{\rm min}(0,0)=0.51,
\]
attained at the forward and backward endpoints.  Thus every
angle-conditioned unpolarized state satisfies $\Neg>1/2$ and witnesses
$\SN\geq3$, whereas $\Neg_{\rm peak}<1$ and the sufficient
Schmidt-number-four witness is not reached for unpolarized beams.

\subsection{Polarization dependence and combined scan}
\label{subsec:joint-results}

Scanning the full physical polarization domain, the central BESIII inputs
give critical polarization magnitudes
\[
 P_t^{\rm crit}\simeq0.65,
 \qquad
 P_L^{\rm crit}\simeq0.57
\]
per beam for the pure transverse and equal-and-opposite longitudinal
configurations, respectively.

Table~\ref{tab:threshold-uncertainty} summarizes the propagated
helicity-parameter uncertainties.  Although the central longitudinal
threshold is lower, its propagated variation is wider.  All 10000 parameter
samples cross the transverse threshold, whereas 5.0\% remain below the
longitudinal witness threshold even at $P_L=1$.  The transverse threshold
is therefore less sensitive to the propagated helicity-parameter
uncertainties within the adopted uncertainty model.

\begin{table*}[t]
\caption{Critical polarizations and their propagated uncertainties from
10000 correlated parameter samples.  Central denotes the final-fit BESIII
parameter values adopted in this analysis.  Medians and central intervals are conditional
on a crossing; the last column gives the fraction without a crossing for
polarization magnitudes up to unity.  The first two rows refer to the
polarization of each beam; the last refers to a single longitudinally
polarized electron beam through Eq.~\eqref{eq:single-beam-mapping}.}
\label{tab:threshold-uncertainty}
\begin{ruledtabular}
\begin{tabular}{lccccc}
Threshold & Central & Median & 68\% interval & 95\% interval
& No crossing (\%)\\
\hline
$P_t^{\rm crit}$   & $0.65$ & $0.65$ & $[0.62,0.68]$ & $[0.58,0.71]$ & $0.0$\\
$P_L^{\rm crit}$   & $0.57$ & $0.57$ & $[0.49,0.68]$ & $[0.43,0.83]$ & $5.0$\\
$|P_e^{\rm crit}|$ & $0.86$ & $0.86$ & $[0.79,0.93]$ & $[0.73,0.98]$ & $5.0$\\
\end{tabular}
\end{ruledtabular}
\end{table*}

\begin{figure*}[!t]
\centering
\includegraphics[width=\textwidth]{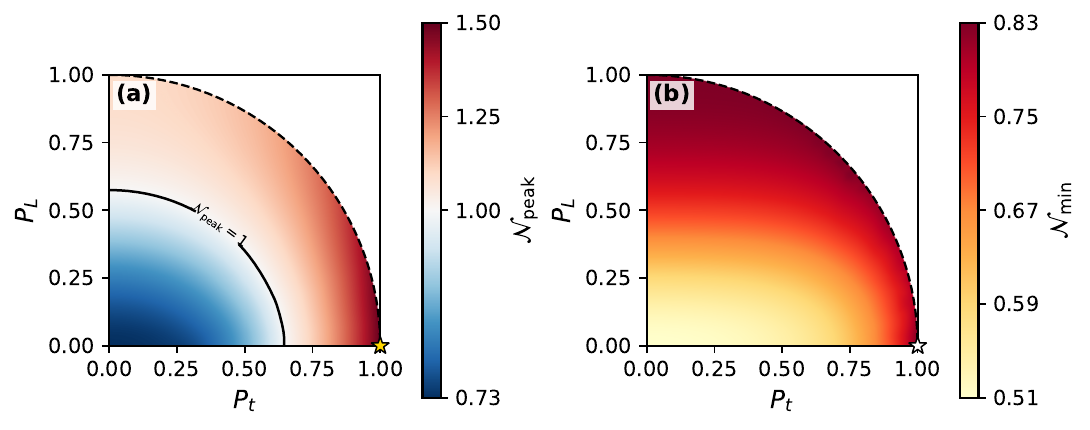}
\caption{Polarization landscape for the BESIII helicity amplitudes.
(a) Angular maximum of the negativity; the black contour marks the
four-dimensional witness threshold $\Neg_{\rm peak}=1$, and the star
denotes the peak optimum $(P_t,P_L)=(1.00,0.00)$.
(b) Angular minimum, with the white star marking the point where $\Neg_{\rm min}$
is maximal, at $(P_t,P_L)=(1.00,0.00)$.  The dashed arc denotes the fully polarized
boundary $P_t^2+P_L^2=1$.}
\label{fig:polarization-landscape}
\end{figure*}

The full polarization landscape is shown in
Fig.~\ref{fig:polarization-landscape}.  Although the longitudinal threshold
is lower, the global maximum of the negativity occurs for pure transverse
polarization.  At the quoted precision,
$\Neg_{\rm peak}\simeq1.50$ at
\[
 (P_t,P_L)=(1.00,0.00),
 \qquad
 \cos\theta=0,
 \qquad
 \phi=\frac{\pi}{2},\frac{3\pi}{2}.
\]
At these symmetry-related angles, the coherent $m=\pm1$ contributions
cancel the $\lambda_1-\lambda_2=\pm1$ channels.  The remaining pure state
contains the four equal-helicity components, with normalized Schmidt
coefficients
\[
 \frac{(h_1,h_1,h_4,h_4)}
 {\sqrt{2(h_1^2+h_4^2)}}.
\]
Its negativity is therefore
\begin{equation}
 \Neg_\star
 =
 \frac32-\frac{(h_1-h_4)^2}{h_1^2+h_4^2}.
 \label{eq:transverse-analytic-negativity}
\end{equation}
The near equality of the measured $h_1$ and $h_4$ explains the
near-maximal four-dimensional entanglement at this point.  The relative
phases do not affect these Schmidt coefficients at this special production
direction, whereas phase-dependent interference away from it shapes the
angular extent of the witness.

For comparison, the angular minimum increases from $0.51$ in the
unpolarized case to values between $0.78$ and $0.83$ along the fully
polarized boundary as the transverse--longitudinal orientation varies.
Its largest value, $0.83$, occurs at the pure-transverse point
$(P_t,P_L)=(1.00,0.00)$.  Thus polarization broadly enhances the
least favorable angle-conditioned spin states, even though the
Schmidt-number-four witness remains confined to more restricted angular
regions.

\begin{figure*}[!t]
\centering
\includegraphics[width=\textwidth]{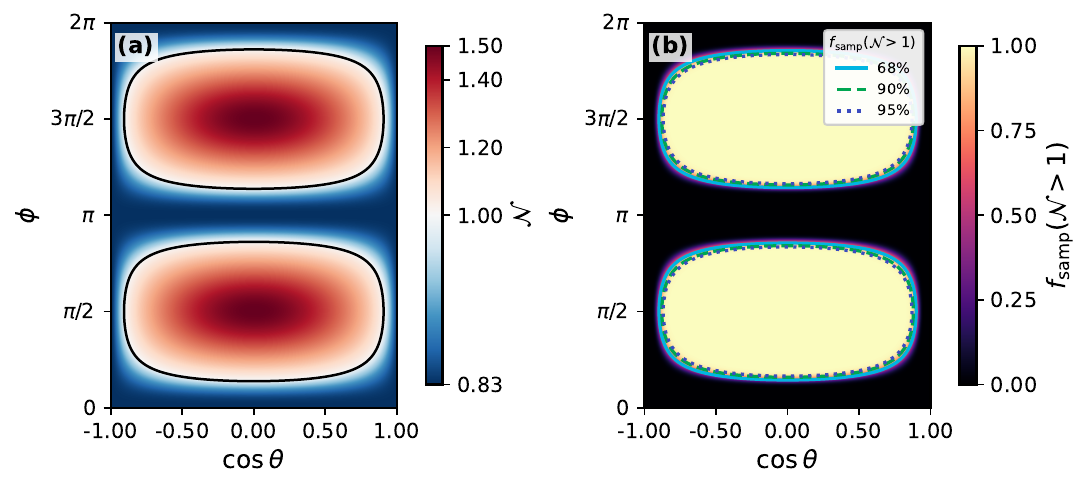}
\caption{Angular structure and uncertainty robustness at the pure-transverse
optimum $(P_t,P_L)=(1.00,0.00)$.  (a) Negativity for the central helicity
parameters, with the black contour marking $\Neg=1$.  (b) Sample fraction
$f_{\rm samp}(\Neg>1)$ of the 10000 correlated BESIII parameter samples
satisfying the witness at each production angle.  The contours indicate
$f_{\rm samp}=68\%$, $90\%$, and $95\%$.}
\label{fig:angular-robustness}
\end{figure*}

At the pure-transverse optimum, the witness forms two symmetry-related
angular lobes centered near
$(\cos\theta,\phi)=(0,\pi/2)$ and $(0,3\pi/2)$, as shown in
Fig.~\ref{fig:angular-robustness}(a).  The production-rate-weighted fraction
defined in Eq.~\eqref{eq:weighted-witness-fraction} is
\[
 f_{\Neg>1}=0.60.
\]
Thus approximately 60\% of the total production rate lies in angular
regions where the pointwise state satisfies the four-dimensional witness
$\Neg>1$.

The propagated BESIII parameter uncertainties preserve the two-lobe witness
structure, as shown by the sample fraction $f_{\rm samp}(\Neg>1)$ defined
in Eq.~\eqref{eq:sample-witness-fraction} and displayed in
Fig.~\ref{fig:angular-robustness}(b).  Using the production rate
corresponding to the central BESIII parameters as the angular weight, the
region with $f_{\rm samp}(\Neg>1)\geq0.95$ accounts for
\[
 f_{95}^{(\rm c)}=0.52
\]
of the production-weighted angular domain, as defined in
Eq.~\eqref{eq:robust-witness-fraction}.

For the sample-specific weighted witness fraction, where each parameter
sample is weighted by its own production rate, the median is $0.60$, with
central 68\% and 95\% intervals of $[0.55,0.64]$ and $[0.50,0.68]$,
respectively.  The peak negativity has median $1.50$, with corresponding
intervals of $[1.49,1.50]$ and $[1.48,1.50]$.  The lower edge of the 95\%
interval for $\Neg_{\rm peak}$ remains well above the four-dimensional
witness threshold, demonstrating the stability of the certification within
the adopted uncertainty model.

\begin{figure*}[!t]
\centering
\includegraphics[width=\textwidth]{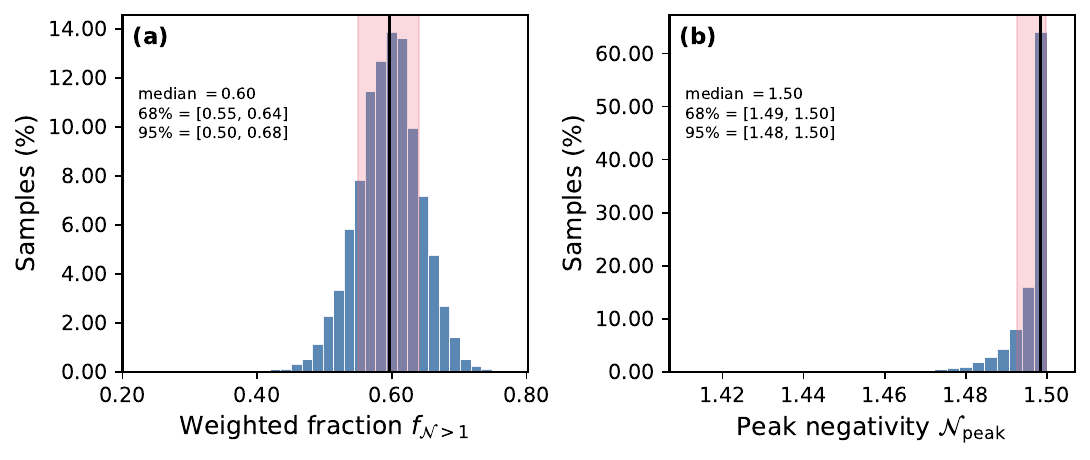}
\caption{Propagated BESIII helicity-parameter uncertainties at the
pure-transverse optimum.  (a) Production-rate-weighted angular fraction
satisfying $\Neg>1$, with each parameter sample weighted by its own
production rate.  (b) Peak negativity.  The vertical black lines mark the
medians, and the darker and lighter shaded bands indicate the central
68\% and 95\% intervals, respectively.  The distributions illustrate the
propagated variation in both the angular extent and the strength of the
four-dimensional witness.}
\label{fig:global-uncertainties}
\end{figure*}

\subsection{Quark-model phase benchmarks}
\label{subsec:quark-results}

The two phase benchmarks share the model magnitudes in
Eq.~\eqref{eq:quark-ratios}, isolating the role of phase-dependent
interference in the polarization response
(Table~\ref{tab:quark-comparison}).

\begin{table}[!tbp]
\caption{Polarization and entanglement observables for the two quark-model
phase benchmarks.  The last row gives the production-rate-weighted angular
fraction satisfying the pointwise condition $\Neg>1$ at the peak-optimal
polarization.  The polarization that maximizes $\Neg_{\rm min}$ need not coincide with
the peak-negativity optimum.}
\label{tab:quark-comparison}
\begin{ruledtabular}
\begin{tabular}{lcc}
Quantity & ${\cal Q}_0$ & ${\cal Q}_{\pi/2}$\\
\hline
Unpolarized $\Neg_{\rm peak}$ & $0.72$ & $0.68$\\
Unpolarized $\Neg_{\rm min}$ & $0.56$ & $0.56$\\
$P_t^{\rm crit}$ & $0.73$ & $0.80$\\
$P_L^{\rm crit}$ & $0.59$ & $0.51$\\
Peak-optimal $(P_t,P_L)$ & $(1.00,0.00)$ & $(1.00,0.00)$\\
Optimized $\Neg_{\rm peak}$ & $1.46$ & $1.40$\\
Maximum $\Neg_{\rm min}$ & $0.92$ & $0.91$\\
Optimal $(P_t,P_L)$ for $\Neg_{\rm min}$ & $(1.00,0.00)$ & $\simeq(0.67,0.74)$\\
$f_{\Neg>1}$ at the peak optimum & $0.76$ & $0.18$\\
\end{tabular}
\end{ruledtabular}
\end{table}

Both benchmarks reach the four-dimensional witness and favor pure
transverse polarization for the peak, as do the BESIII amplitudes.
Changing from ${\cal Q}_0$ to ${\cal Q}_{\pi/2}$ reduces the optimized
peak negativity only from $1.46$ to $1.40$, whereas the weighted witness
fraction decreases from $0.76$ to $0.18$.  Correspondingly,
$P_t^{\rm crit}$ increases from $0.73$ to $0.80$, while
$P_L^{\rm crit}$ decreases from $0.59$ to $0.51$.  The phase assignment
therefore has a substantially larger effect on the angular extent of the
four-dimensional witness than on its optimized peak negativity.

For ${\cal Q}_{\pi/2}$, the angular minimum is instead maximized
on the fully polarized boundary at $(P_t,P_L)\simeq(0.67,0.74)$.  Thus the
polarization configuration maximizing
$\Neg_{\rm peak}$ need not coincide with that maximizing $\Neg_{\rm min}$.

\section{Discussion}
\label{sec:discussion}

\subsection{Physical significance of Schmidt-number-four certification}
\label{subsec:certification-significance}

A spin-$3/2$ baryon spans a four-dimensional local spin space, but the
existence of four local spin states does not by itself imply
four-dimensional entanglement.  The production amplitudes and the
polarization of the parent state determine how these spin degrees of
freedom participate in the bipartite quantum correlations.  For the
angle-conditioned $\Omega^-\bar\Omega^+$ state, $\Neg>1$ is sufficient to
certify Schmidt number four: every pure-state decomposition then requires
at least one component of Schmidt rank four
\cite{TerhalHorodecki2000}.  The certified entanglement therefore requires
the full local spin dimension in a basis-independent sense, rather than
being confined to an effectively two- or three-dimensional entangled
subspace.

For the central measured BESIII helicity parameters, the constructed spin
states witness $\SN\geq3$ throughout the unpolarized angular domain and
are predicted to satisfy the Schmidt-number-four witness in selected angular
regions under suitable beam polarization.  The latter result connects
control of the parent-state populations and coherences directly to the
dimensionality of the produced spin correlations.

The pointwise witness used above refers to an idealized state at fixed
production angles.  For a finite production-angle bin $B$, the corresponding
production state should instead be defined as
\begin{equation}
 \rho_B^{\rm prod}(P_t,P_L)
 =
 \frac{
 \displaystyle
 \int_B d\Omega\,
 R(\theta,\phi;P_t,P_L)
 }{
 \displaystyle
 \Tr\!\int_B d\Omega\,
 R(\theta,\phi;P_t,P_L)
 }.
 \label{eq:finite-bin-production-state}
\end{equation}
The matrices entering this average must be expressed in a common spin
frame.  Since the negativity is nonlinear, the pointwise negativities
cannot in general be averaged directly; the density matrices must first be
combined and the negativity then evaluated for
$\rho_B^{\rm prod}$.  The angular regions identified in this work therefore
provide targets for future polarized-collider measurements through the
cascade-decay angular distribution.  Detector efficiency, angular
resolution, and bin migration must be incorporated into the measurement
and reconstruction model.  Experimental certification would require a
lower confidence bound on
$\Neg(\rho_B^{\rm prod})$ above unity.

\subsection{Why polarization can enhance or suppress entanglement}
\label{subsec:polarization-mechanism}

Longitudinal and transverse polarizations modify the produced spin state in
qualitatively different ways.  Longitudinal polarization changes the
relative populations of the parent $m=+1$ and $m=-1$ components and
therefore reweights their contributions to the final-state helicity
channels.  Transverse polarization of both beams instead introduces
coherence between these components, allowing their decay amplitudes to
interfere.

This distinction explains the complementary behavior found in the
polarization scan.  The longitudinal configuration can reach the
Schmidt-number-four witness at a smaller polarization magnitude per beam,
because population reweighting is already sufficient to move selected
angle-conditioned states across the threshold.  Transverse polarization,
however, can produce a much stronger effect through coherent cancellation
and enhancement of helicity channels.  At the pure-transverse optimum,
the $m=\pm1$ contributions cancel the
$\lambda_1-\lambda_2=\pm1$ channels at
$\cos\theta=0$ and
$\phi=\pi/2,3\pi/2$.  The remaining state is built from the four
equal-helicity components, whose Schmidt coefficients are controlled by
$h_1$ and $h_4$.  Their measured near equality,
$h_1\simeq h_4$, explains why the resulting negativity approaches the
maximum allowed value for a $4\otimes4$ state.

Away from these special production directions, the coherent terms need not
act constructively.  Their sign and magnitude depend on the production
azimuth and on the relative helicity phases, so transverse polarization can
locally enhance or suppress the negativity.  The polarization threshold
and the optimized peak therefore probe different aspects of the dynamics:
the former measures how readily the four-dimensional witness can be
activated, whereas the latter reflects the strongest coherent
reorganization of the helicity channels allowed by the measured
amplitudes.

\subsection{Dependence on the production-amplitude pattern}
\label{subsec:amplitude-pattern}

The illustrative constituent-quark benchmark gives $h_4$ close to the
BESIII central value, with a smaller $h_1$ and a larger $h_3$.  The
zero-phase benchmark ${\cal Q}_0$ shares the pure-transverse peak optimum
of the measured amplitudes and produces a similar optimized negativity.
It also gives a higher angular minimum and a larger
production-rate-weighted witness fraction.  Thus amplitude patterns that
produce comparable peak entanglement can nevertheless generate markedly
different angular structures.

Helicity magnitudes set the relative strengths of the production channels,
while, for fixed magnitudes, the relative phases determine their
interference pattern.  This separation is illustrated directly by the two
quark-model phase benchmarks.  Changing from ${\cal Q}_0$ to
${\cal Q}_{\pi/2}$ leaves the magnitudes unchanged and modifies the
optimized peak negativity only modestly, but changes the
production-rate-weighted angular extent much more strongly.  The peak and
the angular extent therefore provide complementary diagnostics: the peak
is primarily sensitive to whether a favorable combination of helicity
magnitudes can produce a highly entangled state at a particular direction,
whereas the angular extent is more sensitive to how phase-dependent interference
distributes the four-dimensional witness over the full angular domain.

The comparison between ${\cal Q}_0$ and ${\cal Q}_{\pi/2}$ isolates this
phase dependence at fixed model magnitudes.  By contrast, the comparison
between the constituent-quark benchmark and the BESIII amplitudes combines
differences in both magnitudes and phases and should therefore not be
interpreted as a phase-only effect.

\subsection{Implications for STCF and other channels}
\label{subsec:future-implications}

For the central BESIII amplitudes, the transverse witness threshold is
$P_t\simeq0.65$ per beam. Hyperon sensitivity studies have considered
transverse beam polarization generated through the Sokolov--Ternov effect
\cite{Cao2024}. Using Eq.~\eqref{eq:single-beam-mapping}, the
equal-and-opposite longitudinal threshold
$P_L\simeq0.57$ corresponds to
$|P_e|\simeq0.86$ when only the electron beam is longitudinally polarized.
The propagated median is $0.86$, with a conditional 68\% interval of
$[0.79,0.93]$. The benchmark value $|P_e|=0.80$ adopted in an STCF
sensitivity study \cite{Zeng2023} therefore lies within the propagated
interval, although it is below the central single-beam threshold.

The two-beam configuration reduces the polarization magnitude required
per beam because both beams contribute to the longitudinal population
asymmetry of the parent state. The propagated interval also shows,
however, that the precise threshold remains sensitive to the underlying
helicity amplitudes. The polarization values obtained here should
therefore be interpreted as physics benchmarks for entanglement
certification rather than as direct machine-performance requirements.
A realistic experimental sensitivity study would additionally need to
account for the achievable beam polarization and luminosity, together
with detector acceptance, angular resolution, and finite angular binning.

More generally, the same spin-density-matrix framework can be applied to
other spin-$3/2$ baryon-pair channels once the corresponding helicity
amplitudes are available.

\section{Summary and outlook}
\label{sec:summary}

Using the measured BESIII helicity amplitudes, we have shown how beam
polarization enables Schmidt-number-four certification in
$\psitwos\to\OO$. For the central amplitudes, unpolarized production gives
$\Neg=0.51$--$0.73$, certifying $\SN\geq3$ for all production angles.
Polarization activates the sufficient witness $\Neg>1$, with critical
magnitudes of approximately $0.65$ per beam for equal transverse
polarizations and $0.57$ per beam for equal-and-opposite longitudinal
polarizations. Their propagated 68\% intervals are $[0.62,0.68]$ and
$[0.49,0.68]$, respectively, with the longitudinal interval conditional
on the 95.0\% of parameter samples that cross the witness threshold.
Full transverse polarization yields a near-maximal peak
$\Neg_{\rm peak}\simeq1.50$ and a production-rate-weighted angular fraction
of $0.60$ satisfying the pointwise Schmidt-number-four witness.

Longitudinal polarization controls the relative populations of the parent
$m=\pm1$ components, whereas transverse polarization controls their
coherence. At the pure-transverse optimum, destructive interference removes
the unequal-helicity channels at the most favorable production directions;
the measured near equality of $h_1$ and $h_4$ then yields nearly equal
Schmidt coefficients. This mechanism directly connects
polarization-controlled helicity interference to entanglement requiring
the full local spin dimension.

The two illustrative constituent-quark benchmarks share the same model
magnitudes and favor pure transverse polarization for the optimized peak.
Changing the phase assignment from ${\cal Q}_0$ to ${\cal Q}_{\pi/2}$
modifies the peak negativity only modestly, while reducing the
production-rate-weighted witness fraction from $0.76$ to $0.18$.
The angular extent of the witness is therefore substantially more sensitive to the
relative phases than the optimized peak negativity, providing a
phase-sensitive diagnostic complementary to the maximum entanglement
strength.

At full transverse polarization, the propagated 95\% intervals are
$[1.48,1.50]$ for the peak negativity and $[0.50,0.68]$ for the
sample-specific production-rate-weighted witness fraction, supporting the
robustness of the prediction within the adopted uncertainty model.
Experimental certification would require reconstruction of finite-bin
production states in a common spin frame, with detector and angular-resolution
effects incorporated into the measurement model. More broadly, these results
establish beam polarization as a controllable handle on the entanglement
dimensionality of high-spin baryon pairs and provide quantitative benchmarks
for future polarized $e^+e^-$ studies.

\begin{acknowledgments}
We thank the BESIII Collaboration for providing the numerical inputs and
Prof. Jiaojiao Song for helpful discussions.
\end{acknowledgments}

\renewcommand\footnoterule{}
\interlinepenalty=10000
\makeatletter
\renewcommand\bibsection{\section*{References}}
\makeatother
\bibliography{Omegapair}

\appendix
\section{Derivation of the constituent-quark helicity amplitudes}
\label{app:quark-amplitudes}

We summarize the derivation of the constituent-quark helicity amplitudes used in
Sec.~\ref{subsec:quark-input}.  We follow the three-gluon
constituent-quark helicity-amplitude framework of
Ref.~\cite{PangPing2007}; the expressions below give the specialization of
that framework used in this work.  Suppressing color, coupling,
spatial-wave-function, and normalization factors common to all helicity
channels, the spin-dependent Lorentz contraction entering the helicity-amplitude
ratios can be written as
\begin{equation}
\begin{aligned}
 \mathcal K_{\Delta\lambda}
 ={}& (\epsilon_{\Delta\lambda}\!\cdot J_1)(J_2\!\cdot J_3)
 + (\epsilon_{\Delta\lambda}\!\cdot J_2)(J_1\!\cdot J_3)
 \\
 &+ (\epsilon_{\Delta\lambda}\!\cdot J_3)(J_1\!\cdot J_2).
\end{aligned}
 \label{eq:appendix-quark-current}
\end{equation}
where $J_i^\mu=\bar u(p_i)\gamma^\mu v(q_i)$,
$\Delta\lambda\equiv\lambda_1-\lambda_2$, and
$\epsilon_{\Delta\lambda}$ is the polarization four-vector of the parent
vector state.  The momenta $p_i$ and $q_i$ correspond to the $s$ and
$\bar s$ constituents, respectively.

The symmetric spin-$3/2$ wave functions needed for the projection include
\begin{align}
 \chi_{3/2,3/2} &= |\uparrow\uparrow\uparrow\rangle,
 \nonumber\\
 \chi_{3/2,1/2} &=
 \frac{|\uparrow\uparrow\downarrow\rangle
      +|\uparrow\downarrow\uparrow\rangle
      +|\downarrow\uparrow\uparrow\rangle}{\sqrt{3}},
 \label{eq:appendix-spin-wavefunctions}
\end{align}
with the negative-helicity states and the antibaryon states obtained
analogously.  The arrows denote constituent spin projections along the
corresponding baryon momentum.

We evaluate Eq.~\eqref{eq:appendix-quark-current} with canonical Dirac
spinors in the collinear, equal-sharing approximation,
\begin{equation}
 p_i^\mu=\frac{p_\Omega^\mu}{3},
 \qquad
 q_i^\mu=\frac{p_{\bar\Omega}^\mu}{3},
 \qquad
 E_q=\frac{E_\Omega}{3},
 \qquad
 m_q=\frac{M_\Omega}{3}.
 \label{eq:appendix-equal-sharing}
\end{equation}
Choosing the $\Omega^-$ direction as $+z$, the constituent momenta may be
written as $p_i^\mu=(E_q,0,0,k)$ and
$q_i^\mu=(E_q,0,0,-k)$, where
$k=\sqrt{E_q^2-m_q^2}$ is the magnitude of the constituent three-momentum.
Using canonical Dirac spinors in our phase convention gives the spin-resolved
quark currents
\begin{align}
 J^\mu_{\uparrow\uparrow} &= (0,0,0,2m_q),
 \nonumber\\
 J^\mu_{\downarrow\downarrow} &= (0,0,0,-2m_q),
 \nonumber\\
 J^\mu_{\uparrow\downarrow} &= (0,2E_q,-2iE_q,0),
 \nonumber\\
 J^\mu_{\downarrow\uparrow} &= (0,2E_q,2iE_q,0).
 \label{eq:appendix-spin-currents}
\end{align}
Writing the symmetric spin-wave-function coefficients as
$C^{\lambda}_{s_1s_2s_3}$, the helicity projection can be expressed as
\begin{equation}
 \widetilde F_{\lambda_1\lambda_2}
 =
 \sum_{\{s_i,t_i\}}
 C^{\lambda_1}_{s_1s_2s_3}
 C^{\lambda_2}_{t_1t_2t_3}
 \mathcal K_{\lambda_1-\lambda_2}(\{s_i,t_i\}),
 \label{eq:appendix-helicity-projection}
\end{equation}
where the spin labels specify the currents entering
Eq.~\eqref{eq:appendix-quark-current}.  For example, for
$(\lambda_1,\lambda_2)=(3/2,3/2)$ all three constituent spins are aligned,
and the three symmetric contractions contribute equally, giving
\begin{equation}
 \widetilde F_{3/2,3/2}=3(8m_q^3)=24m_q^3.
 \label{eq:appendix-example-channel}
\end{equation}
Evaluating the remaining channels analogously gives, up to the same overall
factor,
\begin{align}
 \widetilde F_{3/2,3/2}
 &=24m_q^3,
 \nonumber\\
 \widetilde F_{3/2,1/2}
 &=-8\sqrt{6}\,E_qm_q^2,
 \nonumber\\
 \widetilde F_{1/2,1/2}
 &=-8m_q\left(3m_q^2-4E_q^2\right),
 \nonumber\\
 \widetilde F_{1/2,-1/2}
 &=-16\sqrt{2}\,E_q\left(2E_q^2-m_q^2\right).
 \label{eq:appendix-helicity-amplitudes}
\end{align}

The signs of individual helicity amplitudes depend on the phase convention
chosen for the external helicity states.  Transforming to the helicity-state
phase convention used in Sec.~\ref{subsec:helicity} reverses the signs of
$\widetilde F_{3/2,1/2}$ and $\widetilde F_{1/2,1/2}$, while leaving the
other two channels unchanged.  Substituting
Eq.~\eqref{eq:appendix-equal-sharing} then reproduces the four helicity amplitudes in
Eq.~\eqref{eq:quark-amplitudes}, up to the common overall factor.  This rephasing is a convention change and
does not alter the physical content.  The common overall factor cancels in
the amplitude ratios used in this work.

A direct numerical evaluation at
$m_{\psi(3686)}=3.686\ \mathrm{GeV}$ and
$M_\Omega=1.6724\ \mathrm{GeV}$ gives
\begin{equation}
 (h_1,h_3,h_4)=(0.4171,0.6061,0.6736),
\end{equation}
which rounds to the values used in Eq.~\eqref{eq:quark-ratios}.

\end{document}